\documentclass[]{pasj01}
\usepackage[switch,mathlines]{lineno}

\jyear{2024}
\Received{}
\Accepted{}
 
\begin{document} 

\title{Evolution of massive black hole in galactic nucleus}

\author{Hajime \textsc{Inoue}\altaffilmark{1}}%
\altaffiltext{1}{Institute of Space and Astronautical Science, Japan Aerospace Exploration Agency, 3-1-1 Yoshinodai, Chuo-ku, Sagamihara, Kanagawa 252-5210, Japan}
\email{inoue-ha@msc.biglobe.ne.jp}



\maketitle

\begin{abstract}
We propose a scenario for mass evolution of massive black holes (MBH) in galactic nuclei, to explain both the mass correlation of the supermassive black hole (SMBH) with the bulge and the down-sizing behavior of the active galactic nuclei.
Primordial gas structures to evolve galactic bulges are supposed to be formed at $z \sim$ 10 and the core region, called the nuclear region (NR) here, is considered to be a place for a MBH to grow to the SMBH.
The down-sizing behavior requires the MBH to significantly increase the mass in a time $\sim$1 Gy.
The rapid mass increase is discussed to be realized only when the MBH stays in a very high density region such as a core of a molecular cloud throughout the period $\sim$1 Gy.
According to these arguments, the MBHs formed from the population III stars born in the mini halos at $z \sim$ 20 – 30 are excluded from the candidates for the seed black hole to the SMBH and only the MBHs from the population II stars born in the core of the central molecular cloud (CMC) in the NR remain as them.
The MBHs in the dense core of the CMC started increasing the mass through mass-accretion and the most massive black hole (MMBH) got the most rapid evolution, possibly restraining relatively slow evolutions of the less massive black holes.
Dynamical interactions of the MMBH with the ambient MCs induced the wandering motion and the further mass-increase.
However, when the MMBH mass exceeded a boundary mass, the dynamical friction with the field stars brakes the MMBH wandering and the mass accretion.
This scenario can semi-quantitatively reproduce both the down-sizing behavior and the SMBH mass - bulge mass correlation with reasonable parameter values.

\end{abstract}


\section{Introduction}\label{Introduction}
It is known that a supermassive black hole (SMBH) generally exist at the center of a galaxy and that the black hole mass well correlates with the mass of the parent galactic bulge (see e.g. Kormendy \& Ho 2014 for the review). 
This discovery of the correlation motivated various theoretical and observational studies, focusing on the co-evolution of the black hole with the ambient galactic environments through several feedback processes
(see e.g. Fabian 2012 for the review).
The majority of the ideas is such that a wind or outflow is driven by the accretion onto the central black hole but that it sweeps away the ambient gas, choking further mass-supply to the black hole.
For example, the simple analytic model on the feedback of the spherical AGN (active galactic nuclei) wind is shown to well reproduce the observed correlation between the SMBH mass and the velocity dispersion of the bulge (King 2003; 2005).
However, recent observations of AGNs with IR and submillimeter interferometers
have revealed that the matter is outflowing in the
polar directions in association with the radiation cones (see,
e.g., H\"{o}nig 2019 and references therein) and those with X-rays support this picture (e.g. Ogawa et al. 2021).  These could indicate that the radiation outflow preferentially distributes around the polar direction, separating from the accretion flow in the equatorial region (see e.g. Inoue 2021b).  It could be difficult for the wind driven by the AGN radiation to give the feedback on the accretion process under such non-spherical  circumstances.

On the other hand, Inoue (2021a) proposed a different idea that the black hole wanders in the galactic nucleus causing the mass accretion, but that dynamical friction by the ambient stars stops the wandering when the black hole mass exceeds a boundary value. However, the quantitative arguments were insufficient there.
The basic idea is succeeded in this paper.

Mass accretion often takes place on to the SMBH, inducing various activities in the galactic nucleus.
Ueda et al. (2003, see also Ueda et al. 2014 for the up to date) obtained the cosmological-redshift-dependent luminosity function of hard X-rays from the AGNs, from which we can approximately calculate the average mass increasing rate of the SMBH.
Exploiting it, Shankar et al. (2004) estimated the present accreted-mass density of AGNs and compared it with the mass density of the SMBH evaluated from the mass correlation between the SMBH and the bulge.
The two mass densities are shown to agree with each other within the errors, from which we can say that the super-mass of the black hole could be the result of the mass accretion originating the AGN activities.

The remarkable finding from the redshift-dependent hard X-ray luminosity function is the cosmic downsizing or anti-hierarchical evolution of the SMBH, indicating that present-day more massive black holes formed in earlier cosmic time than less massive ones (see Ueda 2015 for the review).
This is supported by the observational evidences that very luminous AGNs are preferentially observed as quasars at distances with the cosmological redshift $z$ significantly larger than unity and that hundreds of them have been discovered in the first billion years of cosmic history (see Fan et al. 2023 for the review).
The origin of the downsizing have been discussed by several authors (Marulli et al. 2008; Fanidakis et al. 2012; Hirschmann et al. 2012; 2014, Enoki et al. 2014; Buchner et al. 2015; Delvecchio et al. 2020), but no consensus has been obtained.

In this paper, we study a semi-quantitative scenario on how one of massive black holes (MBH) evolved to the SMBH in a galactic nuclei, to explain both of the mass-correlation between the central black hole and the ambient bulge, and the down-sizing of the AGN activity.

Various studies have been done on formation-pathways of seed black holes and accretion processes to the SMBH (see e.g. Smith \& Bromm 2019; Inayoshi et al. 2020, for the reviews).
The main focus of them is, however, on the formation of $\sim 10^{9} M_{\odot}$ black holes powering the luminous quasars at high cosmological redshifts and little general-discussions have been provided on the mass evolution to the SMBH in the wide mass range from $\sim 10^{6} M_{\odot}$ to $\sim 10^{9} M_{\odot}$ with concerns on the mass correlation and the down-sizing.

The two target relations, the mass correlation and the  down-sizing behavior are briefly introduced in section 2, and a scenario to reproduce them is presented in section 3 with a sequence of the birth of the MBHs from the Pop II stars in the core region of a primordial gas structure to evolve a galactic bulge formed at $z \sim 10$, mass evolution of the most massive  black hole (MMBH) in it, and the final fate of  the MMBH.
Necessary  conditions for the scenario to quantitatively reproduce the two relations  are discussed in  section 4 and the summary and discussion are given in section 5.


\section{Target relations}\label{Targets}

\subsection{Mass correlation between super-massive black hole and the parent bulge}

The relations of the central black hole mass to some properties of the bulge of a parent galaxy have been studied from several observations (see e.g. Kormendy \& Ho 2013 for the review). 
We here assume the correlation between the SMBH mass, $M_{\rm smbh}$ and the bulge mass, $M_{\rm blg}$ as
\begin{equation}
M_{\rm smbh} = 5 \times 10^{-3} M_{\rm blg}
\label{eqn:Msmbh-Mblg}
\end{equation}
for simplicity, although this is slightly different from the precise relation obtained by Kormendy and Ho. 

\subsection{Down-sizing relation}
\begin{figure}
 \begin{center} 
  \includegraphics[width=8cm]{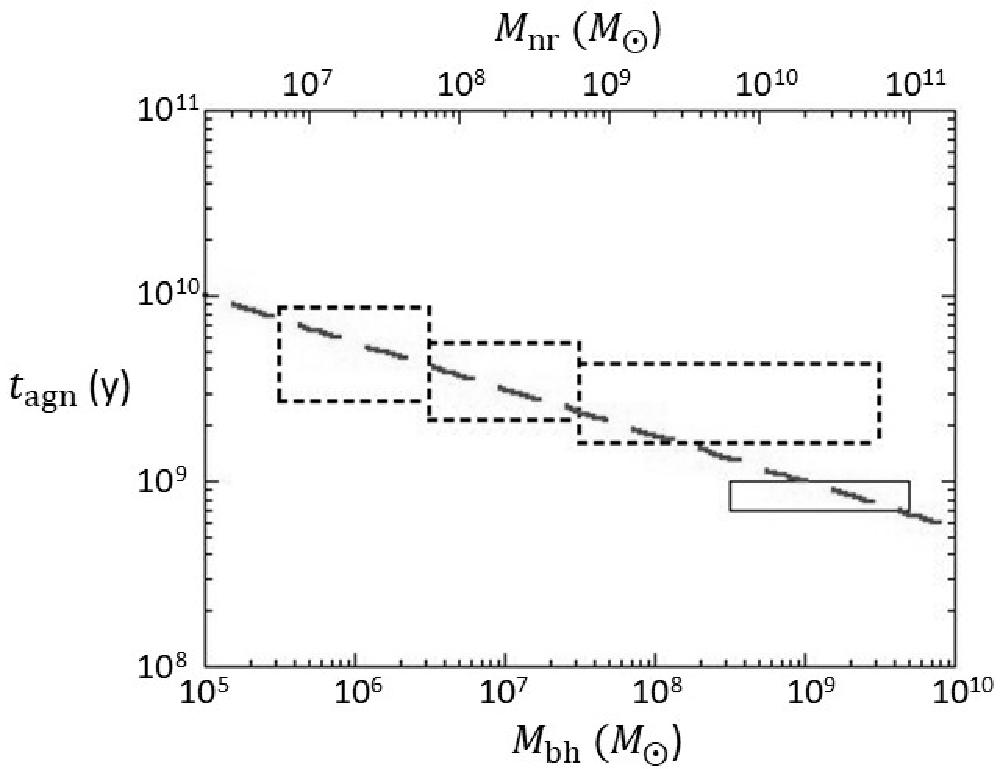} 
 \end{center}
\caption{Relation between the time for the AGN activity to become the highest  in the age of the Universe and the black hole mass, indicated with three domains surrounded by dashed lines (translated from the results by Ueda et al. 2014) and a domain of very distant quasars with thin lines (from Fan et al. 2023).   The top scale shows the nuclear-region mass corresponding to the black hole mass on the bottom scale,  calculated by the relation in equation (\ref{eqn:Msmbh-Mblg}) with help of equation (\ref{eqn:M_nc-M_blg}).  The thick dashed line is an approximate relation given in equation (\ref{eqn:t_agn-M_bh}).  {Alt text: Two line graph.  x axis shows the mass from 10$^5$ to 10$^{10}$ solar mass and y-axis shows the time from 10$^{8}$ to 10$^{11}$ years.} }
\label{fig:DownSizing}
\end{figure}

Figure 12 in Ueda et al. (2014) shows the co-moving space number density of AGNs integrated in different X-ray luminosity, $L_{\rm x}$, bins as a function of cosmological redshift, $z$.
The most active period of the AGNs in each of the luminosity bins can roughly be given by the range of $z$ in which the number density is above the half maximum in the figure, and it is read $\sim$ 0.5 - 2.4,  1.1 - 3.1 and 1.5 - 3.4 for log $L_{\rm x}$ = 43-44, 44-45 and 45-47 respectively.
We see that more luminous AGNs have their number density peak at higher redshifts compared with less luminous ones, indicating the ``downsizing" evolution. 
The luminosity ranges can be transformed to the mass ranges of the black holes responsible for the AGN activities by assuming a relation between the luminosity and the black hole mass, while the age of the Universe when the galactic nuclei with black holes in the respective mass range were active can be calculated from the $z$ values by assuming the cosmological model.
Figure \ref{fig:DownSizing} is the results of such transformations, in which the most active period of the galactic nuclei in the age of the Universe, $t_{\rm agn}$, is plotted as a function of the black hole mass, $M_{\rm bh}$.
For the estimation of the black hole mass, the X-ray luminosity has been assumed to correspond to 10$^{-0.5}$ times the Eddington luminosity, considering that the bolometric luminosity is close to the Eddington luminosity (see e.g. Figure 6 in Fan et al. 2023).
For the cosmological model, the Planck cosmological parameters (Ade et al. 2014) have been used.
In addition, the region of quasars detected at $z > 5.9$ in the $t_{\rm agn}$ -- $M_{\rm bh}$ plane is roughly indicated with a thin line box in the figure, based on Figure 7 of Fan, Banados and Simcoe (2023).

Hereafter, we adopt an approximate relation to represent the overall downsizing trend in figure \ref{fig:DownSizing} as
\begin{equation}
t_{\rm agn} = 10^{9.5} \left( \frac{M_{\rm bh}}{10^{7} M_{\odot}}\right)^{-0.25} \mbox{ years},
\label{eqn:t_agn-M_bh}
\end{equation}
which is shown with a thick dashed line in figure 1.

\section{Scenario for the history of the massive black hole}\label{Scenario}

\subsection{Nuclear Regions formed in the first galaxies}
The recent theoretical studies infer that the formation of the population (Pop) III stars occurred in the so-called mini halo having total mass of $\sim 10^{6} M_{\odot}$ at redshift $z \sim 20 \sim 30$ (see e.g. Bromm 2013; Glover 2013; Klessen \& Glover 2023 for the review).
It is further inferred that the mini-halos were destroyed by the vigorous  negative feedback through strong radiation and supernova explosions of the Pop III stars and that without direct relations to the mini-halos, first galaxies were born at $z \lesssim 10$ through collapses due to radiative cooling of halos with mass $\gtrsim 10^{8} M_{\odot}$ (see e.g. Bromm \& Yoshida 2011 for the review of the first galaxies). 

According to the inference, a self-gravitating isothermal sphere of primordial gas is assumed to be formed with a galactic bulge size when the cosmological redshift, $z$, is around $\sim 10$.
For this size, the contribution of the dark matter to the mass content could be neglected.
This gaseous structure is supposed to evolve a galactic bulge.

The numerical solution of the density profile for the self-gravitating isothermal sphere is shown in Figure 4.6 in Binney and Tremaine (2008) 
and could be divided into two regions: the core with the almost constant density and the envelope with the density decreasing with an approximate dependence on the radial distance, $r$, as $r^{-2}$. 
Thus, we approximate the radial profile of the density, $\rho$, as
\begin{equation}
\rho = \left\{ 
 \begin{array}{ll}
   \rho_{0} &  \mbox{when }r \le R_{\rm c} \\
   \rho_{0}\left(\frac{r}{r_{\rm c}}\right)^{-2}   & \mbox{when } r > R_{\rm c}.
\end{array} \right.
\label{eqn:IsothermalSphere}
\end{equation}
where  $R_{\rm c}$ is the core radius separating the two regions.

Hereafter, we discuss the history of the core region of the primordial gas around the center, which is designated as the nuclear region (NR).
The NR is assumed to be a sphere with a radius, $R_{\rm nr}$ and to have a constant density, $\rho_{\rm nr}$, and a constant average velocity, $\sigma_{\rm nr}$, of the gaseous thermal and turbulent motions.
The total mass, $M_{\rm nr}$, is given as
\begin{equation}
M_{\rm nr} = \frac{4\pi}{3} \rho_{\rm nr} R_{\rm nr}^{3},
\label{eqn:Mnc-rho}
\end{equation}
and the square of the average velocity is calculated from the virial theorem as
\begin{equation}
\sigma_{\rm nr}^{2} \simeq \frac{GM_{\rm nr}}{ R_{\rm nr}},
\label{eqn:sigma_nc}
\end{equation}
where $G$ is the gravitational constant.

The mass of the NR is supposed to be proportional to the total bulge mass, $M_{\rm blg}$ and to be 10$^{9} M_{\odot}$ with $R_{\rm nr} \simeq$ 100 pc for $M_{\rm blg} \simeq 10^{10} M_{\odot}$ considering that the inner region of the nuclear stellar disk of our Galaxy (Launhardt et al. 2002) could be the relic of the NR in this case.
Namely, we have
\begin{equation}
M_{\rm nr} = 10^{-1} M_{\rm blg}.
\label{eqn:M_nc-M_blg}
\end{equation}
We further assume the following relation as
\begin{equation}
R_{\rm nr} = R_{\rm nr, 0} \left( \frac{M_{\rm nr}}{M_{\rm nr, 0}}\right)^{0.5 + \delta},
\label{eqn:R_nc-M_nc}
\end{equation}
where 
\begin{equation}
R_{\rm nr, 0} = 100 \mbox{ pc}, 
\label{eqn:R_nc,0-Def}
\end{equation}
\begin{equation}
M_{\rm nr, 0} = 10^{9} M_{\odot},
\label{eqn:M_nc,0-Def}
\end{equation}
and the value of $\delta$ will be discussed later.

The gas in the NR is expected to be already polluted by metals by supernova explosions of the Pop III stars in the mini-halos and to form such molecular clouds (MC) as observed in the present galaxies.
Here, we suppose that the most of gas in the NR was occupied by a number of discrete MCs giving birth of the Pop  II stars.

\subsection{Seed black holes to the SMBH}\label{SeedBH}
The first candidate for the seed black holes to evolve to the SMBH could be the MBH formed from the Pop III stars in the mini halos.
The cosmological simulations on the first star formation reveal that the Pop III stars have a typical mass around several 10 $\sim$ several 10$^{2} M_{\odot}$ and that a fairly large fraction of the stars could become the massive black hole with several 100 $M_{\odot}$ (Hirano et al. 2014; 2015).
In addition, a possibility for the high mass end to extend up even to $\sim 10^{5} M_{\odot}$ is  discussed by considering the star formation in massive primordial halos with mass more than $10^{7} M_{\odot}$ in which the gas temperature is as high as 10$^{4}$ K (e.g. Toyouchi et al. 2022).

Supposing the NR was formed after the mini halos giving the birth of MBHs were destroyed, it could be considered that the MBHs which happened to be there randomly distribute in the NR.
Then, the mass increase of the MBHs is expected to take place by the accretion  of the ambient gas with the rate which can be approximated as
\begin{equation}
\frac{dM_{\rm bh}}{dt} \simeq  \frac{4\pi (GM_{\rm bh})^{2} \rho_{\rm nr}}{\sigma_{\rm nr}^{3}}.
\label{eqn:Mdot_bh-0}
\end{equation}
Note that no suppression of the accretion rate is considered in this paper, taking into account the significant effect of advection in accretion flow with a super-Eddington rate (see e.g. Inoue 2024). 

Assuming $\rho_{\rm nr}$ and $\sigma_{\rm nr}$ constant in time, the solution of the above equation is obtained as
\begin{equation}
\frac{M_{\rm bh}}{M_{\rm bh,0}} = \left( 1 - \frac{t}{t_{\rm mi}} \right)^{-1},
\label{eqn:M_bh-Evolution}
\end{equation}
where $t$ is time from the start of the mass-increase, $M_{\rm bh, 0}$ is the initial mass and $t_{\rm mi}$ is the mass increase time scale defined as
\begin{equation}
t_{\rm mi, nr} = \frac{\sigma_{\rm nr}^{3}}{4\pi G^{2} \rho_{\rm nr} M_{\rm bh, 0}},
\label{eqn:t_mi-bc}
\end{equation}
in case that the gas fraction in the NR matter is still very high.
If we estimate this time scale for $M_{\rm nr} = 10^{11} M_{\odot}$ and $R_{\rm nr}$ = 1 kpc, it becomes $4 \times 10^{3}$  Gy even for $M_{\rm bh,0} = 10^{5} M_{\odot}$ and is much longer than  the AGN-activity age of $\sim$ 1 Gy for  that  $M_{\rm nr}$ as seen from figure \ref{fig:DownSizing}.
Hence, in order for the MBH to have  the mass increase in such a short time,  a situation could be necessary in which there exists some  place with much higher  density than the average of the NR and the MBH stays there.

The core region of a molecular clound (MC) can provide such a high density  place but a difficult issue is how a MC core traps  a MBH which is randomly moving in the NR.
A possible way could be for the MBH to decrease the kinetic energy through the equipartition with the light Pop II stars which  have already been  born by then and to sink into the MC core around the bottom of the NR potential.
According to section 7.4 in Binney and Tremain (2008), the time scale of the equipartition could be comparable to the relaxation time, $t_{\rm rl}$, of the stellar system in the NR, which is given as
\begin{equation}
t_{\rm rl} \simeq  0.34 \frac{\sigma_{\rm nr}^{3}}{ G^{2} f_{\rm st} \rho_{\rm nr} M_{\rm bh, 0} \ln \Lambda},
\label{eqn:t_rl-bc}
\end{equation}
where $f_{\rm st}$ is the fraction of the stellar mass and $\ln \Lambda$ is a Coulomb logarithm.
Comparing this equation with equation (\ref{eqn:t_mi-bc}) and considering $f_{\rm st} \ll 1$ at the early universe, we see that $t_{\rm rl}$ is also much longer than 1 Gy and thus we cannot expect  for the MBH to  sink into the bottom region of the NR potential. 

These arguments indicate that the  MBHs formed from  the Pop  III stars could be difficult to evolve to the SMBH in the present scenario.

The remaining candidate for the seed black hole  could be  the MBHs formed from Pop II stars born just in the cores of MCs in the NR.
Even in this case, however, the MBH should keep staying in a particular MC core throughout the time scale of $t_{\rm mi} \sim$ 1 Gy to significantly increase the mass, while the MC positions and the situation of the ambient gas to the MBH are  considered to change in a  relatively short time scale as discussed later.
Hence, it could be difficult also for the Pop II MBHs born in the MC cores to evolve to the SMBH in general. 

Only an exception could be MBHs formed in the central MC (CMC) which is expected to exist around the center of the NR.
If we consider the situation in which the space of the NR is occupied by a number of MCs, it could be very natural that such a high density region as the MC core stays at the bottom of the NR potential, since there is no other gravitational source as heavy as it in the early universe.  Thus, we could expect that the MC core keeps staying at the potential bottom until the mass of a MBH exceeds the MC core mass.
Hereafter, we study the history of the Pop II MBHs in the core of the CMC.

\subsection{Mass accretion on to the MBH}

We suppose that the CMC has the mass, $M_{\rm cmc}$ and the boundary radius, $R_{\rm cmc, b}$,  and that it has basically the same structure as in equation (\ref{eqn:IsothermalSphere}) after replacing $\rho_{0}$ and $R_{\rm c}$ by the CMC-core density, $\rho_{\rm cmc, c}$ and the CMC-core radius, $R_{\rm cmc, c}$ respectively.

The mass of the CMC core is expressed as 
\begin{equation}
M_{\rm cmc,c} = \frac{4\pi}{3} \rho_{\rm cmc, c} R_{\rm cmc, c}^{3}, 
\label{eqn:M_cmc}
\end{equation}
while the total mass of the CMC, $M_{\rm cmc}$, can be approximated when $R_{\rm cmc, c} \ll R_{\rm cmc, b}$ as
\begin{equation}
M_{\rm cmc} \simeq 4\pi \rho_{\rm cmc, b} R_{\rm cmc, b}^{3}.
\label{eqn:M_cmc-Cal}
\end{equation}
Here, $\rho_{\rm cmc, b}$ is the CMC density at the boundary which relates to the core density as
\begin{equation}
\rho_{\rm cmc, b} = \rho_{\rm cmc, c} \left(\frac{R_{\rm cmc, c}}{R_{\rm cmc,b}}\right)^{2}.
\label{eqn:rho_b-rho_c}
\end{equation}
Thus, we have a relation as
\begin{equation}
M_{\rm cmc, c} \simeq \frac{1}{3} \frac{R_{\rm cmc,c}}{R_{\rm cmc, b}} M_{\rm cmc}.
\label{eqn:M_cmc,c-M_cmc}
\end{equation}

The mass accretion of the ambient gas in the CMC core can be expected to take place on the MBH there.
The accretion rate is approximately given as in equation (\ref{eqn:Mdot_bh-0}) after replacing $\rho_{\rm nr}$ and $\sigma_{\rm nr}$ with $\rho_{\rm cmc,c}$ and $\sigma_{\rm cmc}$ respectively and the mass-increase time scale, $t_{\rm mi}$, of the MBH in the CMC core is written as
\begin{equation}
t_{\rm mi} = \frac{\sigma_{\rm cmc}^{3}}{4\pi G^{2} \rho_{\rm cmc,c} M_{\rm bh, 0}}.
\label{eqn:t_mi-Def}
\end{equation}
Here, the average velocity of the gas particles in the CMC, $\sigma_{\rm cmc}$, which is assumed to be constant in the whole volume of the CMC, is obtained as
\begin{equation}
\sigma_{\rm cmc}^{2} = \frac{\rho_{\rm nr}}{\rho_{\rm cmc, b} }  \sigma_{\rm nr}^{2},
\label{eqn:sigma_cmc-sigma_nc}
\end{equation}
from the pressure balance of the CMC gas with the NR gas at the boundary, 
where almost all the matter of the NR is assumed to be still gas in this epoch of the very early Universe.

Equation (\ref{eqn:t_mi-Def}) tells us that the rapider evolution of the mass is obtained for the larger $M_{\rm bh,0}$.
Furthermore, the mass increase to the super-massive stage takes place in a very short time at the last moment of the period of $t_{\rm mi}$ as seen from equation (\ref{eqn:M_bh-Evolution}).
Thus, it is expected that the most massive black hole (MMBH) among the MBHs in the CMC core first evolves to the super-massive black hole (SMBH) but that the other MBHs with the smaller masses still remain in the mass range near the initial ones when the MMBH has evolved to the SMBH.
If the AGN activities due to the mass increase of the MMBH disturb the environments of the mass accretion onto the less massive MBHs, the mass evolution of them could be suppressed.
In addition, the Bondi radius of the mass accretion by the MBH, $R_{\rm B}$, which is estimated with help of equations (\ref{eqn:sigma_cmc-sigma_nc}), (\ref{eqn:Mnc-rho}), (\ref{eqn:M_cmc-Cal}), (\ref{eqn:sigma_nc}) and (\ref{eqn:R_nc-M_nc}) as
\begin{eqnarray}
R_{\rm B} &\simeq& \frac{GM_{\rm bh}}{\sigma_{\rm cmc}^{2}} \nonumber \\
&\simeq& \left(\frac{M_{\rm cmc}}{M_{\rm nr,0}}\right)^{2} \left(\frac{R_{\rm nr,0}}{R_{\rm cmc,b}}\right)^{4} \left(\frac{M_{\rm nr}}{M_{\rm nr,0}}\right)^{4\delta} \left(\frac{M_{\rm bh}}{M_{\rm cmc}}\right) \frac{R_{\rm cmc,b}}{3} \nonumber \\
&\simeq& \left(\frac{M_{\rm nr}}{M_{\rm nr,0}}\right)^{1/6} \left(\frac{M_{\rm bh}}{M_{\rm cmc}}\right) \frac{R_{\rm cmc,b}}{3}
\label{eqn:R_B}
\end{eqnarray}
for $M_{\rm cmc} \sim 10^{7} M_{\odot}$, $R_{\rm cmc,b} \sim 10$ pc, $\delta \simeq 1/24$ (adoptions of these parameter values are discussed in subsection \ref{DownSizingReproduction}), $M_{\rm nr,0}$ in equation (\ref{eqn:M_nc,0-Def}) and $R_{\rm nr,0}$ in equation (\ref{eqn:R_nc,0-Def}), 
becomes as large as the size of the CMC core, supposed to roughly be a tenth of $R_{\rm cmc, b}$, when $M_{\rm bh}$ gets close to the CMC mass.  
Hence, all the gas in the CMC core to be accreted by the MBHs could eventually be monopolized by the most massive one which has first increased the mass close to the CMC mass.  
From these arguments, only the MMBH could be considered to have a chance to become the SMBH.
In fact, the most AGNs look consistent to have a single SMBH observationally.

Hereafter, we assume that the MMBH can alone evolve to the SMBH.

\subsection{Wandering of the MMBH around the center of the NR}\label{MMBH-wandering}
The mass of the MMBH is expected to increase due to the mass accretion of the ambient gas in the CMC.
However, the gas content in the CMC is limited and mass supply to the CMC could be needed for the MMBH mass to become as large as $M_{\rm cmc}$ or larger.
Interactions of the CMC with the ambient MCs could be the origin of the further gas supply.

The MCs are considered to basically float supported by the pressure gradient in the gravitational field of the NR.
However, star formation activities disturb the spacial distribution of the MCs in  the NR by consuming the gas mass through the star formation itself, changing the thermal situations through radiation from young stars, inducing gas-flows by the stellar winds or  supernova explosions and so on. 
Thus, the MC positions are expected to change in time in  the NR.

The dynamical interactions of the CMC with the nearby MCs could make it possible for the CMC gas to be supplied and for the MMBH mass to increase even up to a mass more than the original CMC mass.
If the MMBH mass gets larger than the CMC mass, the situation changes from one in which the CMC containing the MMBH interacts with the nearby MCs to another in which the MMBH accompanied by the gas and stellar cluster wanders the central space occupied by the MCs.

Because of the discrete distribution of the MCs, the MMBH receives attractive force mainly from the innermost MC even at the center of the NR.
Let us suppose a situation in which the distribution of nearby MCs varies and the direction to the nearest MC seen from the MMBH among the nearby MCs changes every an average time interval, $t_{\rm  mc}$.
Then, the MMBH would get  the gravitational force from the nearest  MC, $F_{\rm gf}$, which can be approximated as
\begin{equation}
F_{\rm gf} \simeq M_{\rm bh} \frac{G M_{\rm mc}}{R_{\rm nr}^{2}},
\label{eqn:F_gf}
\end{equation}
where $M_{\rm mc}$ is the MC mass and $R_{\rm nr}$ is the distance of the nearest MC from the  MMBH.
The velocity towards the nearest MC, $V$, is roughly calculated as
\begin{equation}
V \sim \frac{GM_{\rm mc}}{R_{\rm nr}^{2}} t_{\rm mc}.
\label{eqn:V-Cal}
\end{equation}

If the MMBH moves due to the gravitational attraction by the nearest MC, the motion could be braked by the dynamical friction by the ambient field stars.
The dynamical friction force on the MMBH from the field stars, $F_{\rm df}$,  is approximately given, referring to section 8.1 of Binnay and Tremaine (2008), as
\begin{equation}
F_{\rm df} \simeq \frac{4\pi (GM_{\rm bh})^{2} \rho_{\rm st, nr} \ln \Lambda}{2 \sigma_{\rm nr}^{2}} \frac{A(X)}{X^{2}},
\label{eqn:F_df-0}
\end{equation}
where $\rho_{\rm st, nr}$ and $\sigma_{\rm nr}$ are the average density of the field stars and the standard deviation of the velocity distribution, respectively, and
\begin{equation}
A(X) = \rm{erf}(X) -\frac{2X}{\sqrt{\pi}}\rm{e}^{-X^{2}}.
\label{eqn:A(X)}
\end{equation}
Here, erf($X$) is the error function and $X$ is defined as
\begin{equation}
X = V/(\sqrt{2}\sigma_{\rm nr}).
\label{eqn:X-Def}
\end{equation}
Since $X$ is much less than unity in the presently considered situation,  $A(X)$ can be approximated as
\begin{equation}
A(X) \simeq 0.75 \ X^{3}.
\label{eqn:A(X)_X<<1}
\end{equation}
$\rho_{\rm st, nc}$ is expressed as
\begin{equation}
\rho_{\rm st, nr} = f_{\rm st} \frac{3 M_{\rm nr}}{4 \pi R_{\rm nr}^{3}},
\label{eqn:rho_st,nc}
\end{equation}
where $f_{\rm st}$ is the fraction of the stellar mass in the total NR mass,
and $\sigma_{\rm nr}$ is given in equation (\ref{eqn:sigma_nc}).
Then, equation (\ref{eqn:F_df-0}) is rewritten also with help of equations (\ref{eqn:V-Cal}) and (\ref{eqn:F_gf}) as
\begin{equation}
F_{\rm df} \simeq F_{\rm gf}  \frac{ 2.25 f_{\rm st} \ln \Lambda}{2^{3/2}} \frac{M_{\rm bh}}{M_{\rm nr}} \frac{t_{\rm mc}}{t_{\rm nr}},
\label{eqn:F_df-1}
\end{equation}
where $t_{\rm nr}$ is the dynamical time scale of the NR defined as
\begin{equation}
t_{\rm nr} \equiv \frac{R_{\rm nr}}{\sigma_{\rm nr}},
\label{eqn:t_bc-Def}
\end{equation}
and is estimated as
\begin{equation}
t_{\rm nr} \simeq 4..7 \times 10^{5} \frac{M_{\rm nr}}{M_{\rm nr,0}}^{0.25+1.5\delta} \mbox{ y,}
\label{eqn:t_bc-Cal}
\end{equation}
for the relation in equation (\ref{eqn:R_nc-M_nc}) and the values in equations (\ref{eqn:R_nc,0-Def}) and  (\ref{eqn:M_nc,0-Def}).

We see from equations (\ref{eqn:F_gf}) and (\ref{eqn:F_df-0}) that $F_{\rm gf} \propto M_{\rm bh}$ while $F_{\rm df} \propto M_{\rm bh}^{2}$.
This indicates that there exists a boundary MMBH mass, $M_{\rm bh, b}$,  satisfying $F_{\rm gf} = F_{\rm df}$, and that the stronger one between $F_{\rm gf}$ and $F_{\rm df}$ is determined by the larger one between $M_{\rm bh}$ and $M_{\rm bh, b}$.
When $M_{\rm bh} < M_{\rm bh, b}$, the gravitational force dominates the dynamical friction force, while, when $M_{\rm bh} > M_{\rm bh, b}$, the dynamical friction prevails.

For the mass-increase history of the MMBH, let us  consider the case when $M_{\rm bh} < M_{\rm bh,b}$ first.
In this case, we can expect a situation in which the MMBH almost freely wanders in a region with a radius of $V \times t_{\rm mc}$.
This wandering motion could induce the mass accretion on to the MMBH and the mass could get to the boundary mass, $M_{\rm bh, b}$.

Once $M_{\rm bh}$ exceeds the boundary mass, the  dynamical friction is considered to become dominant to the gravitational attraction from the nearest MC, forcing  the MMBH to stop  wandering.

\subsection{Final mass of the MMBH}\label{MMBH-FinalMass}

Most studies of accretion environment of AGNs performed so far have been based on the picture that a SMBH sits at the center of a galactic nucleus and the accreted matter is supplied from the outside of the nucleus (see e.g. Jogee 2006; Alexander \& Hickox 2012, for the 
review and references therein).
However, the processes for matter to inflow from a region with a $\sim$kpc distance to a region with a 10 $\sim$ 100 pc distance have fairly well been studied both observationally 
and theoretically, but the final path from the 10 - 100 pc region to an accretion-disk  region with a sub-pc distance to the black hole is still far from understanding, facing  difficult issues of the angular momentum barrier and the star-burst barrier (see Jogee 2006; Alexander \& Hickox 2012).

Here, we rather admit the presence of such barriers and assume that the gas supply from the outside cannot proceed inward beyond the 10 - 100 pc region.  
In this case, if $M_{\rm bh}$ exceeds $M_{\rm bh,b}$ and the wandering range gets smaller than the 10 - 100 pc region, it is expected that  the accretion on to the MMBH stops and the mass increase of the MMBH comes to end with the  final mass slightly larger than $M_{\rm bh,b}$. 

$M_{\rm  bh, b}$ is obtained by  equating $F_{\rm gf}$ in equation (\ref{eqn:F_gf}) to $F_{\rm df}$ in equation (\ref{eqn:F_df-1}) as
\begin{equation}
M_{\rm bh, b} = B M_{\rm nr},
\label{eqn:M_bh,b-M_nc-S}
\end{equation}
where
\begin{equation}
B \equiv \frac{2^{3/2}}{2.25 \ln \Lambda} \frac{1}{f_{\rm st}} \frac{t_{\rm nr}}{t_{\rm mc} }.
\label{eqn:B-Def}
\end{equation}
This can be approximately regarded as the final mass of the MMBH.


\section{Reproduction of the two target relations}\label{Reproduction}
Now, we try to reproduce the two relations summarized in section \ref{Targets} with the scenario presented in section \ref{Scenario}, by tuning the relevant parameter values.

\subsection{Down-sizing relation}\label{DownSizingReproduction}
First, we replace the approximate expression of the down-sizing relation in equation (\ref{eqn:t_agn-M_bh}), with help of equations (\ref{eqn:Msmbh-Mblg}) and (\ref{eqn:M_nc-M_blg}), as
\begin{equation}
t_{\rm mi} \simeq 2 \times 10^{9} \left(\frac{M_{\rm nr}}{ M_{\rm nr,0}}\right)^{-0.25} \mbox{ y},
\label{eqn:t_mi-M_nc}
\end{equation}
assuming $t_{\rm agn} \simeq t_{\rm mi}$.

On the other hand, $t_{\rm mi}$ in equation (\ref{eqn:t_mi-Def}) is rewritten, with help of equations (\ref{eqn:sigma_cmc-sigma_nc}), (\ref{eqn:sigma_nc}), (\ref{eqn:Mnc-rho}), (\ref{eqn:rho_b-rho_c}), (\ref{eqn:M_cmc-Cal}) and (\ref{eqn:R_nc-M_nc}), as
\begin{equation}
t_{\rm mi} \simeq \frac{3^{3/2}}{G^{1/2}} \frac{1}{M_{\rm bh,0}} \frac{R_{\rm cmc, b}^{15/2}}{M_{\rm cmc}^{5/2}} \left(\frac{R_{\rm cmc,c}}{R_{\rm cmc,b}}\right)^{2} \frac{M_{\rm nr, 0}^{3}}{R_{\rm nr,0}^{6}} \left(\frac{M_{\rm nr}}{M_{\rm nr, 0}}\right)^{-6\delta}.
\label{eqn:t_mi-cmc,c}
\end{equation}

\begin{figure}
 \begin{center} 
  \includegraphics[width=8cm]{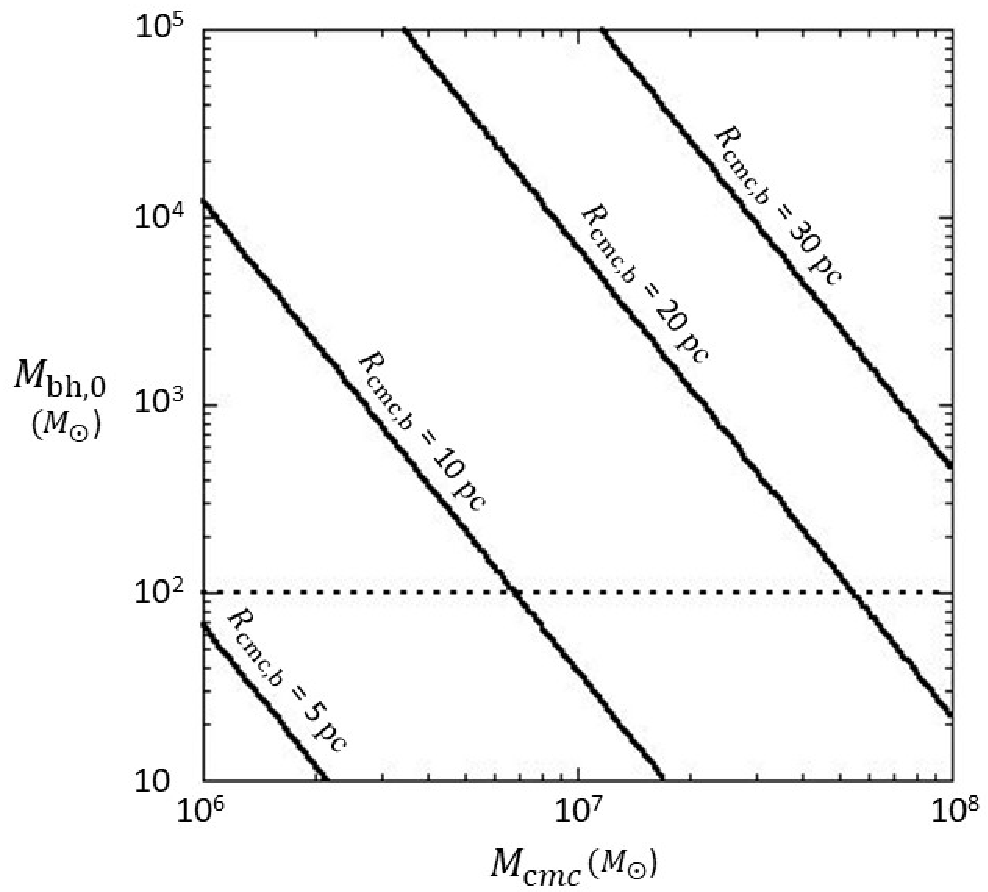} 
 \end{center}
\caption{Relations of $M_{\rm bh, 0}$ to $M_{\rm cmc}$ for $R_{\rm cmc, b}$ = 5, 10, 20 and 30 pc.  The dotted line for $M_{\rm bh,0} = 10^{2} M_{\odot}$ roughly indicates the upper limit of the black hole mass born from the Pop II stars.  {Alt text: Two line graph.  x axis shows the mass from 10$^{6}$ to 10$^{8}$ solar mass and y-axis shows the mass from 10 to 10$^{5}$ solar mass.}}
\label{fig:M_cmc-M_bh0}
\end{figure}

For $t_{\rm mi}$ in equation (\ref{eqn:t_mi-cmc,c}) to satisfy the down-sizing relation given in equation (\ref{eqn:t_mi-M_nc}), the following two relations are necessary as
\begin{equation}
\delta = 1/24,
\label{eqn:delta-Cal}
\end{equation}
to reproduce the dependence on $M_{\rm nr}$
and
\begin{equation}
M_{\rm bh, 0} = 3.9 \times 10^{1} \left(\frac{M_{\rm cmc}}{10^{7} M_{\odot}}\right)^{-5/2}\left(\frac{R_{\rm cmc, b}}{10 \ \rm{pc}}\right)^{15/2} \ M_{\odot},
\label{eqn:M_bh,0-Cal}
\end{equation}
to reproduce the proportional constant in equation (\ref{eqn:t_mi-M_nc}), adopting $R_{\rm cmc, c} = 0.1\ R_{\rm cmc, b}$, $M_{\rm nr, 0}$ in equation (\ref{eqn:M_nc,0-Def}) and $R_{\rm nr,0}$ in equation (\ref{eqn:R_nc,0-Def}).
The relations of $M_{\rm bh, 0}$ to $M_{\rm cmc}$ in cases of $R_{\rm cmc, b}$ = 5, 10, 20 and 30 pc are plotted in figure \ref{fig:M_cmc-M_bh0}.
As seen from the results, if $M_{\rm cmc} \sim 10^{7} M_{\odot}$ and $R_{\rm cmc, b} \sim$ 10 pc, the initial mass of the MMBH can be as small as several 10 $M_{\odot}$ to reproduce the down sizing relation, which is possible to be formed from the Pop II stars.

\subsection{Correlation between the SMBH mass and the NR mass}\label{MassCorrelationReproduction}
In the scenario presented in section \ref{Scenario}, the final mas of the MMBH is given in equations (\ref{eqn:M_bh,b-M_nc-S}) and (\ref{eqn:B-Def}).
This final mass can be considered to correspond to the SMBH mass adopted in the observed relation in equation (\ref{eqn:Msmbh-Mblg}), if the present scenario can reproduce the mass correlation between the SMBH and the bulge.
In that case, $B$ defined in equation (\ref{eqn:B-Def}) should satisfy the following equation, taking account of equation (\ref{eqn:M_nc-M_blg}), as
\begin{equation}
B \simeq 5 \times 10^{-2},
\label{eqn:B-Const}
\end{equation}
over the entire range of $M_{\rm nr}$ considered here.

In order to estimate the $B$ value in equation (\ref{eqn:B-Def}), we have to evaluate the values of $f_{\rm st}$ and $t_{\rm mc}$.

As discussed in subsection \ref{MMBH-wandering}, $t_{\rm mc}$ is the time scale of the direction-change of the  nearest MC viewed from the MMBH and could be the results of the star-formation activities in the MC system of the NR.  Hence, it could be  reasonable to consider that $t_{\rm mc}$ is proportional to the  time scale of the star formation in the NR, $t_{\rm sf}$, and to introduce a relation as
\begin{equation}
t_{\rm mc} = \alpha \ t_{\rm sf},
\label{eqn:t_mc-t_sf}
\end{equation}
where $\alpha$ is the proportional constant.

To evaluate the  $f_{\rm st}$ and  $t_{\rm sf}$ values, we discuss the MC distribution in the NR and the star formation history there.

\subsubsection{MC distribution in the  NR}\label{MC-Distribution}
It is known that the number distribution of the MCs can be approximated by a form of $dN_{\rm mc}/dM_{\rm mc} \propto M_{\rm mc}^{\gamma}$ with the cutoff at the uppermost mass and $\gamma$ is 1.5 $\sim$ 1.7 (see the reviews by Heyer \& Dame 2015 for MCs in our Galaxy; Fukui \& Kawamura 2010 for those in nearby galaxies).
If the mass ranges from $M_{\rm mc, min}$ to $M_{\rm mc, max}$, the average MC mass, $\bar{M}_{\rm mc, 0}$, is calculated to be approximately $(M_{\rm min}/M_{\rm max})^{0.5} M_{\rm max}$ for $M_{\rm max} \gg M_{\rm min}$ and $\gamma = 1.5$.
According to the mass distributions of the MCs in our Galaxy (Heyer \& Dame 2015) and the nearby galaxies (Fukui and Kawamura 2010), the maximum masses are 10$^{6} \sim 10^{6.5} M_{\odot}$ and have no apparent dependence on the mass of the host galaxy.
On the other hand, the distributions extend down to $10^{4} M_{\odot}$.
Thus, we approximate that
\begin{equation}
\bar{M}_{\rm mc, 0} = 10^{5} M_{\odot},
\label{eqn:M_mc,0-Def}
\end{equation}
setting $M_{\rm max} \simeq  10^{6} M_{\odot}$ and $M_{\rm min} \simeq  10^{4} M_{\odot}$ irrespectively of  the mass of the NR.

Then, the total mass of MCs in the NR, $M_{\rm mc, t}$, is expressed with the average mass as
\begin{equation}
M_{\rm mc, t} = N_{\rm mc} \bar{M}_{\rm mc,0},
\label{eqn:M_mc,t}
\end{equation}
where $N_{\rm mc}$ is the total number of the MCs in the NR.
$M_{\rm mc, t}$ is also written as
\begin{equation}
M_{\rm mc,t} \simeq f_{\rm gs} M_{\rm nr},
\label{eqn:Int-f_gs}
\end{equation}
by considering that almost all the gas mass exists in the MCs, and introducing the fraction of the gas mass in the total NR mass, $f_{\rm gs}$. 

Here, we introduce a concept of the one-MC sphere in which  only one MC exists on average and approximate the radius, $R_{1}$, as
\begin{equation}
R_{1} \simeq N_{\rm mc}^{1/3} R_{\rm nr}.
\label{eqn:R_1-R_nc}
\end{equation}
Then, $N_{\rm mc}$ changes in proportion to $M_{\rm mc, t}$ as seen from equation (\ref{eqn:M_mc,t}), and the size of the region occupied by one MC, $R_{1}$, given in equation (\ref{eqn:R_1-R_nc}), decreases as $M_{\rm mc, t}$ increases.
On the other hand, we know that the size distribution of the MCs in our Galaxy have the peak around 30 pc (Miville-Deschenes et al. 2017).
The peak positions are similar to each other between the clouds in the inner Galaxy and those in the outer Galaxy (Miville-Deschenes et al. 2017) and the MCs even in the Galactic center region have the sizes around 30 pc (Oka et al. 1998).
Furthermore, the sizes of the MCs in the nearby galaxies also distribute around 30 pc (see figure 3 in Fukui \& Kawamura 2010).
Thus, we can consider that the average radius is always $R_{\rm mc,0}$ given as
\begin{equation}
R_{\rm mc,0} = 30 \ \rm{pc},
\label{eqn:R_mc,0}
\end{equation}
irrespectively of the environments.

From the above arguments, we introduce a constraint on $R_{1}$ as 
\begin{equation}
R_{1} \ge R_{\rm mc,0}, 
\label{eqn:R1>Rmc0}
\end{equation}
and it is translated to a constraint on $N_{\rm mc}$ with the help of equation (\ref{eqn:R_1-R_nc}) as
\begin{equation}
N_{\rm mc} \le N_{\rm mc, 1},
\label{eqn:N_mc<=N_mc1}
\end{equation}
where
\begin{eqnarray}
N_{\rm mc, 1} &\equiv& \left(\frac{R_{\rm nr}}{R_{\rm mc,0}}\right)^{3} \nonumber \\
&=& \left(\frac{R_{\rm nr,0}}{R_{\rm mc,0}}\right)^{3} \left(\frac{M_{\rm nr}}{M_{\rm nr,0}}\right)^{1.5+3\delta}
\label{eqn:N_mc1}
\end{eqnarray}
with the help of equation (\ref{eqn:R_nc-M_nc}).
Then, we assume the following equation as
\begin{equation}
N_{\rm mc} = \mbox{Min} (N_{\rm mc,1}, N_{\rm mc, 2}),
\label{eqn:N_mc-1&2}
\end{equation}
where
\begin{eqnarray}
N_{\rm mc, 2} &\equiv& \frac{M_{\rm mc, t}}{\bar{M}_{\rm mc, 0}} \nonumber \\
&=& f_{\rm gs} \frac{M_{\rm nr}}{\bar{M}_{\rm mc, 0}}.
\label{eqn:N_mc2}
\end{eqnarray}

\begin{figure}
 \begin{center} 
  \includegraphics[width=8cm]{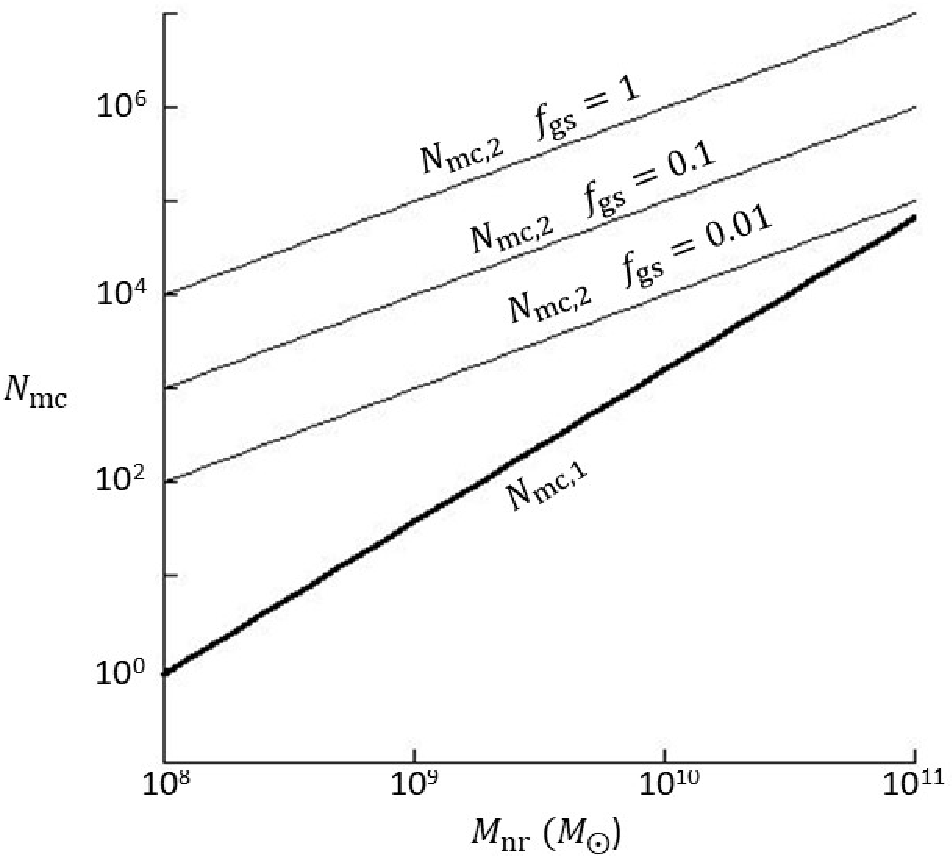} 
 \end{center}
\caption{Plots of $N_{\rm mc,1}$ vs. $M_{\rm nr}$ (thick line) and $N_{\rm mc,2}$ vs. $M_{\rm nr}$ in the case of $\bar{M}_{\rm mc,0} = 10^{5} M_{\odot}$ (thin lines) for $f_{\rm gs}$ =1 (top), 0.1 (middle) and 0.01 (bottom).    {Alt text: Two line graph.  x axis shows the mass from 10$^{8}$ to 10$^{11}$ solar mass and y-axis shows the numerical value from 0.1 to 10$^{7}$.} }
\label{fig:N_mc}
\end{figure}

Figure \ref{fig:N_mc} plots the $N_{\rm mc, 1}$ and $N_{\rm mc, 2}$ distributions as a function of $M_{\rm nr}$.  For $N_{\rm mc, 2}$, three cases of $f_{\rm gs}$ = 1, 0.1 and 0.01 are shown for $\bar{M}_{\rm mc,0} = 10^{5} M_{\odot}$.

On this assumption, when $N_{\rm mc,2} < N_{\rm mc,1}$, the MCs sparsely distributes and the mass is around the typical one, $\bar{M}_{\rm mc, 0}$.
On the other hand, when $N_{\rm mc,2} > N_{\rm mc,1}$, the whole volume is fully occupied by the typical MC size and $N_{\rm mc}$ saturates at $N_{\rm mc, 1}$. In this case, the average mass of MCs, $\bar{M}_{\rm mc,1}$, is calculated as
\begin{eqnarray}
\bar{M}_{\rm mc,1} &=& \frac{f_{\rm gs} M_{\rm nr}}{N_{\rm mc, 1}} \nonumber \\ 
&\simeq& 2.7 \times 10^{7} f_{\rm gs} \left(\frac{M_{\rm nr}}{M_{\rm nr,0}}\right)^{-(0.5+3\delta)} \ M_{\odot},
\label{eqn:M_mc-N_mc1}
\end{eqnarray}
and is larger than the average mass, $\bar{M}_{\rm mc, 0}$, in the case of $N_{\rm mc,2}$, probably due to cloud-cloud merging. 

\subsubsection{Star formation history}\label{StarFormationHistory}
We assume that the star formation rate per unit volume is expressed as 
\begin{equation}
\frac{d\rho_{\rm st}}{dt} = C \rho_{\rm gs}^{\beta}
\label{eqn:SF-Rate}
\end{equation}
where $\rho_{\rm st}$ and $\rho_{\rm gs}$ are the densities of the stellar and gaseous matter respectively, and $C$ is the proportional constant.

Let $f_{\rm gs}$ and $f_{\rm st} = 1-f_{\rm st}$ be the fractions of the gaseous and stellar mass respectively, then the above equation for the star formation rate can be rewritten as
\begin{equation}
\frac{df_{\rm gs}}{dt} = - \frac{f_{\rm gs}^{\beta}}{t_{\rm sf}}
\label{eqn:Eq-f_gs}
\end{equation}
where $t_{\rm sf}$ is the time scale of the star formation defined as
\begin{equation}
t_{\rm sf} \equiv C^{-1} \rho_{\rm tot}^{-(\beta-1)}.
\label{eqn:t_s}
\end{equation}
Here, $\rho_{\rm tot} = \rho_{\rm st} + \rho_{\rm gs}$, and is assumed to be constant in time.
The solution of this equation for $\beta > 1$ is
\begin{equation}
f_{\rm gs} = \left( 1 + \frac{t}{t_{\rm sf}} \right)^{-1/(\beta-1)},
\label{eqn:f_gs-Sol}
\end{equation}
in case that $f_{\rm gs} = 1$ at $t=0$.

If the gas fraction, $f_{\rm gs, p}$,  at the present time, $t_{\rm p}$, the time scale, $t_{\rm sf}$, can be estimated as
\begin{equation}
t_{\rm sf} = \frac{1}{f_{\rm gs, p}^{-(\beta-1)} -1} t_{\rm p}.
\label{eqn:t_s-Est}
\end{equation}

In the NSD in out Galaxy, $f_{\rm gs, p}$ is about 0.5 \% (Launhardt et al. 2002) at the present age of the Universe, $t_{\rm p} \simeq 1.4 \times 10^{10}$ y considering that the star formation history started in the very early universe.  We adopt these values to the NR in case of $M_{\rm nr} = M_{\rm nr,0} = 10^{9} M_{\odot}$, and designate $t_{\rm sf}$ calculated with them as $t_{\rm sf, 0}$ for a given $\beta$.

The star  formation is considered to take place in the MCs. 
In the case of $N_{\rm mc} = N_{\rm mc,2}$, the average mass of the MCs is always $\bar{M}_{\rm mc,0}$ irrespectively of the NR mass.  If we assume that the average radius of the MCs is also the same over the NR mass range as considered here, the average density and thus $t_{\rm sf}$ become independent of $M_{\rm nr}$.
However, in the case  of $N_{\rm mc} = N_{\rm mc,1}$, $\rho_{\rm tot}$  in equation  (\ref{eqn:t_s}) is considered to be proportional to $M_{\rm nr}/R_{\rm nr}^{3}$.
Then, $t_{\rm sf}$ is represented from equation (\ref{eqn:t_s}) by using a relation of $\rho_{\rm tot} \propto M_{\rm nr}/R_{\rm nr}^{3} \propto (M_{\rm nr}/M_{\rm nr,0})^{-(0.5+3\delta)}$ as
\begin{equation}
t_{\rm sf} = t_{\rm sf,0} \left(\frac{M_{\rm nr}}{M_{\rm nr,0}}\right)^{(0.5+3\delta)(\beta-1)},
\label{eqn:t_s-M_nc/M_nc,0}
\end{equation}
for a given $\beta$, where $t_{\rm sf,0}$ is the value in case of $M_{\rm nr}=M_{\rm nr,0}$.

\subsubsection{Best parameters to reproduce the mass correlation}\label{BestParameters}
Now, we look for the best parameters for $B$  in equation (\ref{eqn:B-Def}) to reproduce the empirical relation in equation (\ref{eqn:B-Const}).

\begin{figure}
 \begin{center} 
  \includegraphics[width=8cm]{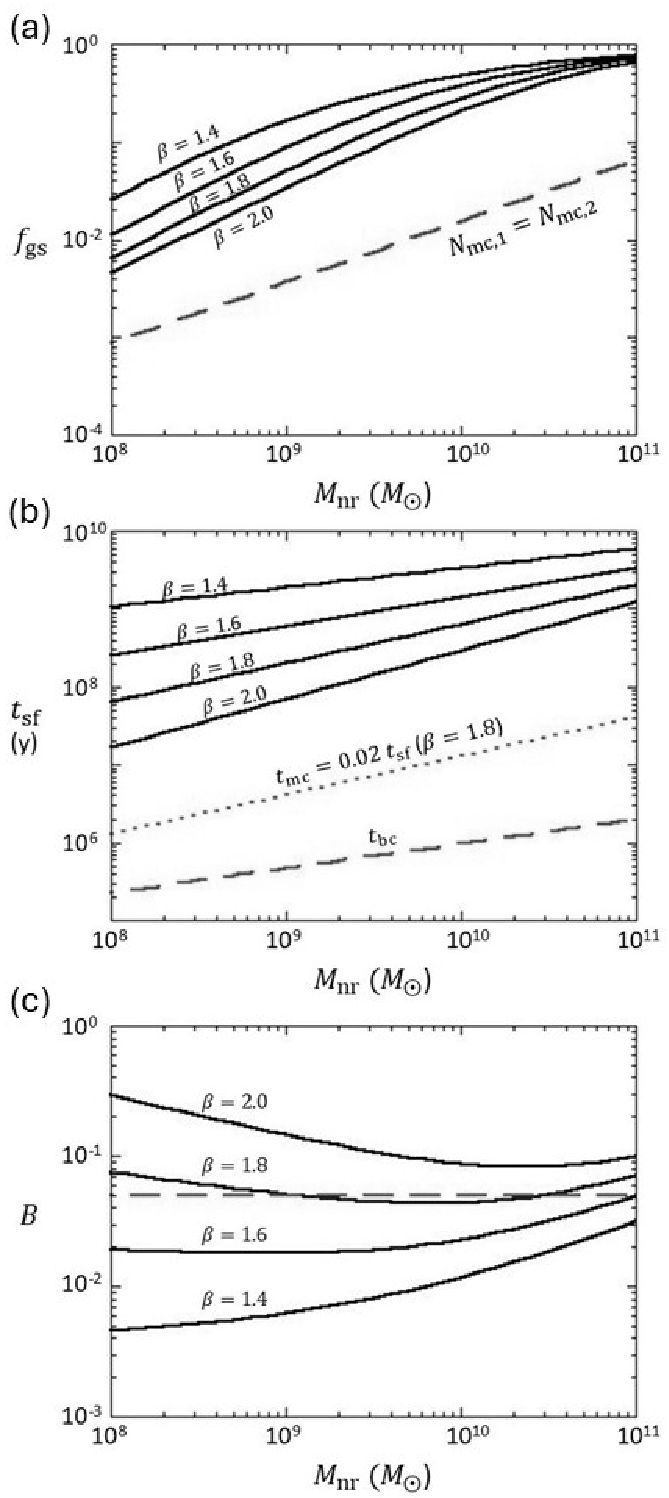} 
 \end{center}
\caption{Plots of (a) the gas fraction of the NR, $f_{\rm  gs}$, at the age of $t_{\rm mi}$ given in equation (\ref{eqn:t_mi-M_nc}) against $M_{\rm nr}$ ;  (b) the star formation time scale, $t_{\rm sf}$, against $M_{\rm nr}$  ; (c)  the parameter $B$ estimated from equation (\ref{eqn:B-Def}) against $M_{\rm nr}$.  Four thick lines in each panel represent the respective values for $\beta$ = 1.4, 1.6, 1.8 and 2.0.  
A dashed line in each of the panel  indicates (a) the  $f_{\rm gs}$ value with which $N_{\rm mc, 1} = N_{\rm mc, 2}$ is satisfied; (b) the dynamical time scale as a function of $M_{\rm nr}$; (c) the proportional constant between $M_{\rm bh}$ and $M_{\rm nr}$ obtained from observations, given  in equation (\ref{eqn:B-Const}).  
A dotted line in the panel (b) presents the time scale of the MC-distribution change in the NR, $t_{\rm mc}$, in case of $t_{\rm mc} = 0.02 t_{\rm sf} (\beta=1.8)$.
These results are obtained by assuming $N_{\rm mc} = N_{\rm mc,1}$.  The presence of the four thick lines above the  $N_{\rm mc,1} = N_{\rm mc,2}$ line in the panel (a) shows the consistency of the calculated $f_{\rm gs}$ values  with the assumption.    {Alt text: Two line graphs.  In the top panel, x axis shows the mass from 10$^{8}$ to 10$^{11}$ solar mass and y-axis shows the numerical value from 10$^{-4}$ to 1.  In the middle panel, x axis shows the mass from 10$^{8}$ to 10$^{11}$ solar mass and y-axis shows the time from 10$^{5}$ to 10$^{10}$ years.  In the bottom panel, x axis shows the mass from 10$^{8}$ to 10$^{11}$ solar mass and y-axis shows the numerical value from 10$^{-3}$ to 1.} }
\label{fig:Mbh-Mnc-Cor}
\end{figure}

In the case of $N_{\rm mc} = N_{\rm mc,1}$, $t_{\rm sf}$ is estimated by equation (\ref{eqn:t_s-M_nc/M_nc,0}) for a given $\beta$.
On the other hand, since we are now considering the final fate of the MMBH in the mass increase history, $f_{\rm gs}$ and $f_{\rm st}$ should be calculated by setting $t = t_{\rm mi}$ which is given by equation (\ref{eqn:t_mi-M_nc}).
Then, we can estimate $f_{\rm gs}$ and $f_{\rm st}$ as a function of $M_{\rm nr}$ for a given $\beta$ from equation (\ref{eqn:f_gs-Sol}).
The results on $f_{\rm gs}$ in four cases of $\beta$ = 1.4, 1.6, 1.8, and 2.0, and $\delta$ in equation (\ref{eqn:delta-Cal}) are plotted in figure \ref{fig:Mbh-Mnc-Cor}-(a). 
$\beta = 1.4$ is known as the typical value through a number of studies of star formation phenomena (see e.g. McKee \& Ostriker 2007).
In this figure, the $f_{\rm gs}$ values when $N_{\rm mc,1} = N_{\rm mc, 2}$ are indicated with the dashed line, the domain above which corresponds to the case of $N_{\rm mc} = N_{\rm mc,1}$.
This figure assures the consistency of the calculated $f_{\rm gs}$ values with the premise of  $N_{\rm  mc} = N_{\rm mc,1}$

Figure \ref{fig:Mbh-Mnc-Cor}-(c) shows the value of $B$ as a function of $M_{\rm nr}$ calculated from equation (\ref{eqn:B-Def}) with the results of $g_{\rm st}$ and $t_{\rm sf}$.  Here, we assume $\alpha = 0.02$ and adopt $\ln \Lambda \sim 3$.  
$\Lambda$ is defined to be  $b_{\rm max}/b_{90}$ in the chapter 8 of Binney and Tremaine (2008), where $b_{90}$ is approximated to be $GM_{\rm bh}/\sigma_{\rm nr}^{2}$ and $b_{\rm max} \sim R_{\rm nr}$ in the present case, indicating $\Lambda \sim M_{\rm nr}/M_{\rm bh}$ with help of equation (\ref{eqn:sigma_nc}) and then $\ln \Lambda \sim 3$ is obtained for $M_{\rm bh}/M_{\rm nr} \sim 5 \times 10^{-2}$ at the final fate of the MMBH. 

As seen from figure \ref{fig:Mbh-Mnc-Cor}-(c), the calculated line for $\beta$ = 1.8 fairly well reproduce the proportional constant between the SMBH mass and the NR mass obtained from observations.
The proportional constant, $\alpha$, of $t_{\rm mc}$ to $t_{\rm sf}$ defined in equation (\ref{eqn:t_mc-t_sf}) has been adjusted to 0.02 to obtain the well fitted case with $\beta = 1.8$.
The time scale, $t_{\rm mc}$, in case of $\alpha$ = 0.02 and $\beta$ = 1.8 is plotted against $M_{\rm nr}$ with a dotted line in figure \ref{fig:Mbh-Mnc-Cor}-(b).
As seen from this figure, the time scale of the MC-distribution change in the NR, $t_{\rm mc}$, in this plot is about 10 $\sim$ 20 times as long as the dynamical time scale of the NR, $t_{\rm nr}$, which could be possible to be realized.

\begin{figure}
 \begin{center} 
  \includegraphics[width=8cm]{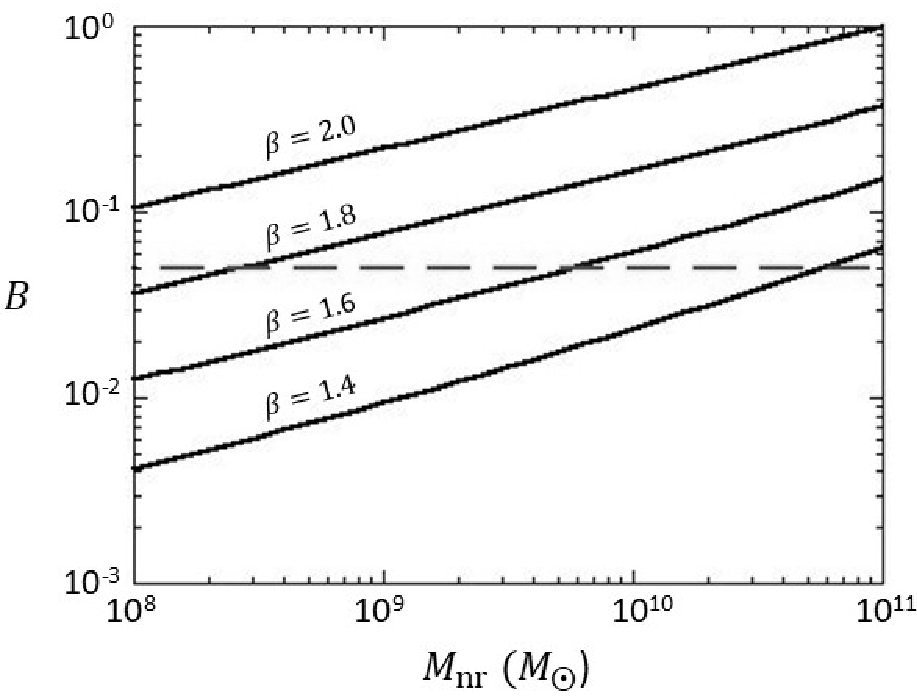} 
 \end{center}
\caption{The plot of $B$ against $M_{\rm nr}$ in the similar manner as in figure \ref{fig:Mbh-Mnc-Cor} -(c), but the case is $N_{\rm mc} = N_{\rm mc,2}$. {Alt text: Two line graphs.  x axis shows the mass from 10$^{8}$ to 10$^{11}$ solar mass and y-axis shows the numerical value from 10$^{-3}$ to 1.}}
\label{fig:Mbh-Mnc-Cor-2}
\end{figure}

On the other hand, if we remove the constraint in equation (\ref{eqn:R1>Rmc0}), 
the situation becomes the case of $N_{\rm mc}=N_{\rm mc,2}$.  Then, $B$ in equation (\ref{eqn:B-Def}) is calculated,  assuming that  every MC has the same density  and thus the same $t_{\rm sf}$ irrespectively of $M_{\rm nr}$, which  is shown in figure \ref{fig:Mbh-Mnc-Cor-2}.   This reveals that the observed relation cannot be reproduced with this case.

\section{Summary and discussion}\label{Discussion}
As shown in the previous section, the target two relations presented in section \ref{Targets} can be reproduced with allowable parameter values in the scenario discussed in section \ref{Scenario}.

We have supposed that self-gravitating isothermal spheres of primordial gas which would evolve to galactic bulges were formed at $z \sim$ 10 and that the core region, the NR, was the stage for the mass evolution of the MBH.
Since the final MBH mass comes to have the correlation with the bulge mass as the result of the mass increase, the overall mass distribution over the core region and the outer region in the proto-bulge structure should not significantly change during a few Gyr of the mass increase (AGN) period in the early Universe as seen in figure \ref{fig:DownSizing}.

\subsection{Reproduction of the down-sizing relation}
The down sizing relation of the AGN activities means that the active period of galactic nuclei with the more massive SMBH came the earlier in the history of the Universe, which can be reworded, because of the mass correlation between the SMBH and the bulge, as
the mass-increase to the SMBH in the more massive NR took the shorter time.

Equations (\ref{eqn:M_bh-Evolution}) and (\ref{eqn:t_mi-bc}) approximately show that the mass-increase time scale of the MBH is determined by the initial mass and the environments soon after the birth in the early universe. 

It has been discussed in subsection \ref{SeedBH} that the high density environment as in the core of a MC is needed for the MBH to  have such a short mass-increase time scale as $\sim$1 Gy and that the MBH to evolve to the SMBH is required to stay there over the mass increase time scale.
These discussions have induced an inference that the MBHs formed from the Pop III stars born in the mini halos cannot be the seed BH to the SMBH and the only a candidate could be the MBHs generated from the Pop II stars just born in the core of the CMC, as far as we consider the growth path of the MBH as presently studied.

Here, the Pop II MBHs born in the MCs other than the CMC have also been excluded from the candidates of the seed black holes.  It is because those MBHs have been considered to be impossible to keep  staying in the cores over the entire time of the mass-increase $\sim$ 1 Gy under the situation in which the gas distribution in the form of the MC system around the MBHs changes more rapidly than the mass-increase time.
In this study, the time scale of the MC-distribution change, $t_{\rm mc}$, has been adjusted to reproduce the mass correlation between the SMBH and the bulge.  The result is shown in figure \ref{fig:Mbh-Mnc-Cor} - (b), which is consistent with the situation of $t_{\rm mc} \ll t_{\rm mi}$.

Then, the CMC has been treated as an exception in that it could keep staying around the center of the NR irrespectively of the MC distribution change.
It is because the most massive object around the center is considered to make the bottom of the NR potential and to have been the CMC in the early universe until the mass of the MMBH exceeds the CMC mass.
In fact, the nuclear star clusters (NSC,  see Neumayer et al. 2020 for the review) are often observed in galactic nuclei, which could be the relic of the CMC.
It should be noted that the fraction of early type galaxies with the NSC is observed to decrease as the galaxy mass increases beyond $\sim 10^{10} M_{\odot}$ (see Fig.3 in Neumayer et al. 2020).
This might be due to the situation change of the CMC caused by the mass increase of the MMBH above the CMC mass. 

In subsection \ref{DownSizingReproduction}, we have seen that if the mass and the size of the CMC are $\sim 10^{7} M_{\odot}$ and $\sim$10 pc respectively, the initial mass of the MBH can be smaller than $\sim 10^{2} M_{\odot}$ to reproduce the down-sizing relation, which fits the mass range required for the seed black holes born from Pop II stars in the CMC.
These values are consistent with the mass and the size distributions obtained from observations of the NSCs (Neumayer et al. 2020).

The number density of gas in the core of the CMC with the mass of $10^{7} M_{\odot}$ and the boundary radius of 10 pc is calculated with equations (\ref{eqn:M_cmc-Cal}) and (\ref{eqn:rho_b-rho_c}) to be $\sim 3 \times 10^{8}$ cm$^{-3}$ assuming $R_{\rm cmc,b}/R_{\rm cmc,c}$ = 10.
This density is very high, responding to the requirement of such a short mass-increase time as $\sim$1 Gy.

The average velocity of the gas particles in the CMC, $\sigma_{\rm cmc}$, is another environmental parameter than the density to determine the mass-increase time scale as seen in equation (\ref{eqn:t_mi-Def}).
$\sigma_{\rm cmc}$ is obtained with equation (\ref{eqn:sigma_cmc-sigma_nc}) considering the pressure balance of the CMC gas with the ambient NR gas at the boundary, and the term of $\rho_{\rm nr} \sigma_{\rm nr}^{2}$ in the right hand side of this equation introduces the dependence of the mass-increase time on $M_{\rm nr}$ as $M_{\rm nr}^{-\delta}$ in equation (\ref{eqn:t_mi-cmc,c}).

The negative power of the dependence of $t_{\rm agn}$ on $M_{\rm bh}$ in the down-sizing relation represented with equation (\ref{eqn:t_agn-M_bh}) is determined by the condition on $\delta$ as given in equation (\ref{eqn:delta-Cal}).
According to the virial relation in equation (\ref{eqn:sigma_nc}), the $R_{\rm nr}$ - $M_{\rm nr}$ relation can be translated to the $M_{\rm nr}$- $\sigma_{\rm nr}$ relation as
\begin{equation}
M_{\rm nr} \propto \sigma_{\rm nr}^{4/(1-2 \delta)}.
\label{eqn:M_nc-sigma_nc-Rel}
\end{equation}
The power of the above equation becomes 4.4 for the value of $\delta$ given in equation (\ref{eqn:delta-Cal}).
This is roughly consistent with the relation between the galaxy total stellar mass, $M_{\rm gal}$, and the velocity dispersion, $\sigma_{\rm e}$, as $M_{\rm gal} \propto \sigma_{\rm e}^{4.7}$ obtained from early type galaxies on the high mass side by Cappellari et al. (2013), and also with the $M_{\rm blg}$ - $\sigma_{\rm e}$ relation derived from the $M_{\rm bh}$ - $M_{\rm blg}$ ralation and the $M_{\rm bh}$ - $\sigma_{\rm e}$ relation in Kormendy and Ho (2013).

\subsection{Reproduction of the mass correlation between the SMBH and the bulge}

The mass increase of the MMBH has been considered to start through the accretion of the ambient gas with the turbulent velocity of $\sigma_{\rm cmc}$ in the CMC core.
If the accreted gas is limited within the gas in the CMC, however, it should be impossible for the MMBH mas to exceed the total gas mass of the CMC.
Then, we have discussed a situation in which positions of the MCs could change in the NR and interactions of the CMC with the neighboring  MCs could supply the additional gas to the CMC, allowing the MMBH to increase the mass even beyond the CMC mass.

If the MMBH mass exceeds the CMC mass, the issue would change to the  situation of the MMBH motion in the central region of the NR, which has been discussed in subsection \ref{MMBH-wandering}.
It has  been indicated that the MMBH could wander  when the mass is smaller than the boundary mass, $M_{\rm  bh, b}$ but that the dynamical friction from the field stars prevent the wandering  when the mass exceeds the boundary mass.
As the result, we could expect that the mass increase of the MMBH could stop at the mass slightly larger  than $M_{\rm bh, b}$.

$M_{\rm bh, b}$ is derived by the balance between the gravitational force on the MMBH from the nearest MC, $F_{\rm gf}$, and the dynamical friction force from the field star, $F_{\rm df}$, as discussed in subsection \ref{MMBH-wandering}.
$F_{\rm df}$ is formulated with the density of the field star and the moving velocity of the MMBH against the field star system as in equation (\ref{eqn:F_df-0}) and can be rewritten as in equation (\ref{eqn:F_df-1}) using equations (\ref{eqn:V-Cal}) and (\ref{eqn:rho_st,nc}) and introducing $t_{\rm nr}$ defined in equation (\ref{eqn:t_bc-Def}).
Then, we have obtained the final mass of the MMBH in the form of equation (\ref{eqn:M_bh,b-M_nc-S}) with the parameter $B$ defined in equation (\ref{eqn:B-Def}).

In order for the final MMBH mass, $M_{\rm bh, b}$, in equation (\ref{eqn:M_bh,b-M_nc-S}) to fit the target mass-correlation between the SMBH and the bulge, the suitable parameters for $B$ to satisfy the condition given in equation (\ref{eqn:B-Const}) have been searched in subsection \ref{MassCorrelationReproduction}.
The key parameters are the fraction of the stellar mass in the NR, $f_{\rm st}$, at the moment when the MMBH mass increased close to the final one, $M_{\rm bh, b}$, and the time scale of star formation in the NR, $t_{\rm sf}$, under the assumption that the time scale of the direction-change of the nearest MC, $t_{\rm mc}$, is proportional to $t_{\rm sf}$ as given in equation (\ref{eqn:t_mc-t_sf}).
To estimate $t_{\rm sf}$ and $f_{\rm st}$, the distribution of the MC in the NR and the star formation history there have been discussed in subsubsections \ref{MC-Distribution} and \ref{StarFormationHistory}. 
Since we need the average quantities over the NR space and the MMBH mass-increase time, the average values of the mass and the size of MCs in the NR have been adopted in the estimations.

The parameter search for $B$ in equation (\ref{eqn:B-Def}) to fit the constant value of $B$ against $M_{\rm nr}$ in equation (\ref{eqn:B-Const}) to reproduce the observed mass relation between the SMBH and the bulge has been done in subsubsection \ref{BestParameters}.
It  has been found that $B$ calculated with $\beta$ = 1.8 in the case of $N_{\rm mc} = N_{\rm mc, 1}$ for $R_{\rm mc,0}$ = 30 pc can well simulate the constant relation.
$\beta$ = 1.8 is consistent with the $N_{\rm  mc, 1}$ case in which  the higher star formation activity than the usual situation with $\beta \simeq$ 1.4 could be expected because of the crowded situation of MCs.

It should be noted that in the case of $N_{\rm mc} = N_{\rm mc, 1}$, the number of MC in the NR becomes $\sim$1 when $M_{\rm nr} = 10^{8} M_{\odot}$ as seen from figure \ref{fig:N_mc}.
Thus, we cannot expect the interactions between the CMC and the ambient MCs which induce the mass increase of the MMBH above the CMC mass, in the lowest range of the NR mass considered 
here.
The discussions to reproduce the mass correlation between the SMBH and the bulge in the present scenario cannot be applied to the mass range of $M_{\rm nr} \lesssim 10^{8} M_{\odot}$.

\subsection{Peculiarities of the CMC}

As discussed above, we have found a possible solution to interpret the mass correlation between SMBH in the situation that the NR space is fully occupied with MCs with the average radius of 30 pc and that the average MC mass is determined with equation (\ref{eqn:M_mc-N_mc1}).

On the other hand, the radius of $\sim$10 pc and the mass of $\sim 10^{7} M_{\odot}$ are required  to the CMC irrespectively of the NR mass to explain the down-sizing relation in the present scenario.
These values deviate from the above average values but the CMC as a member of the MC system could be possible to have them by chance in the mass-radius probability distribution. 
However, it would be unlikely for the randomly selected values to be always the same to the systems with different NR masses.
Furthermore, the mass and radius are required to be kept the same over the almost whole period of the mass increase time scale $\sim$1 Gy, while the MC distribution on a NR is considered to change on a time scale of the order of 10 My as discussed in subsubsection \ref{BestParameters}.

This could be reconciled by considering the peculiar situation of the CMC.  
As discussed before, the CMC should be kept staying at the bottom of the gravitational potential of the NR until the MMBH mass comes to exceed the CMC mass, because there is no other gravitational source as heavy as it around the center in the early universe. 
Also, the mass inflow from the outside to the central region could be weak because of the angular momentum barrier, which could be another reason for the exception. 

One observational evidence of the peculiarity could be that the NSC and/or the SMBH are observed instead of the MC in the central region of present galaxies, which could be regarded as the results of the peculiar history of the CMC.
However, why and how the mass and the radius are selected irrespectively of the NR mass in the CMC is open to future study.

As shown above, the scenario presented in this paper can well reproduce the two target relations.
The study, however, depends on some disputable premises and assumptions, and is still on a level of  order estimation.
More detailed studies will  obviously  be needed.

\begin{ack}
The author is grateful to the referee for the critical and helpful comments.
\end{ack}



\end{document}